\newcommand{\myname}{Geoff Boeing}
\newcommand{\myemail}{boeing@usc.edu}
\newcommand{\myaffiliation}{Department of Urban Planning and Spatial Analysis\\Sol Price School of Public Policy\\University of Southern California}
\newcommand{\paperdate}{August 2020}
\newcommand{\papertitle}{The Right Tools for the Job: The Case for Spatial Science Tool-Building}
\newcommand{\papercitation}{Boeing, G. 2020. \papertitle. \emph{Transactions in GIS}, published online ahead of print. \href{https://doi.org/10.1111/tgis.12678}{doi:10.1111/tgis.12678}}
\newcommand{\paperkeywords}{Urban Planning, Transportation, Data Science}

\RequirePackage[l2tabu,orthodox]{nag}     
\documentclass[11pt,letterpaper]{article} 

\usepackage[T1]{fontenc}                
\usepackage[utf8]{inputenc}             
\usepackage{crimson}                    
\usepackage{helvet}                     

\usepackage[strict,autostyle]{csquotes} 
\usepackage[USenglish]{babel}           
\usepackage{microtype}                  

\usepackage{apacite}

\newcommand{\doi}[1]{https://doi.org/#1}
\usepackage{ragged2e}

\usepackage{abstract}                   
\usepackage{authblk}                    
\usepackage{booktabs}                   
\usepackage{caption}                    
\usepackage[final]{draftwatermark}      
\usepackage{endnotes}                   
\usepackage{geometry}                   
\usepackage{graphicx}                   
\usepackage{hyperref}                   
\usepackage{natbib}                     
\usepackage{rotating}                   
\usepackage{setspace}                   
\usepackage{titlesec}                   
\usepackage{url}                        

\graphicspath{{./figures/}}

\titleformat{\section}{\normalfont\sffamily\large\bfseries\color{black}}{\thesection.}{0.3em}{}
\titleformat{\subsection}{\normalfont\sffamily\small\bfseries\color{black}}{\thesubsection.}{0.3em}{}

\hypersetup{
pdfauthor={\myname},
pdftitle={\papertitle},
pdfsubject={\papertitle},
pdfkeywords={\paperkeywords},
pdffitwindow=true,         
breaklinks=true,           
colorlinks=false,          
pdfborder={0 0 0}          
}

\SetWatermarkText{DRAFT}
\SetWatermarkScale{1.3}
\SetWatermarkLightness{0.9}

\begin{document}

\renewcommand{\APACrefYearMonthDay}[3]{\APACrefYear{#1}}

\title{The Right Tools for the Job:\\The Case for Spatial Science Tool-Building}
\author[]{\myname}
\affil[]{\myaffiliation}
\date{\paperdate}

\maketitle

\begin{abstract}
This paper was presented as the 8\textsuperscript{th} annual \textit{Transactions in GIS} plenary address at the American Association of Geographers annual meeting in Washington, DC. The spatial sciences have recently seen growing calls for more accessible software and tools that better embody geographic science and theory. Urban spatial network science offers one clear opportunity: from multiple perspectives, tools to model and analyze nonplanar urban spatial networks have traditionally been inaccessible, atheoretical, or otherwise limiting. This paper reflects on this state of the field. Then it discusses the motivation, experience, and outcomes of developing OSMnx, a tool intended to help address this. Next it reviews this tool's use in the recent multidisciplinary spatial network science literature to highlight upstream and downstream benefits of open‐source software development. Tool-building is an essential but poorly incentivized component of academic geography and social science more broadly. To conduct better science, we need to build better tools. The paper concludes with paths forward, emphasizing open-source software and reusable computational data science beyond mere reproducibility and replicability.\footnote{{Preprint of: \papercitation. Email: \href{mailto:\myemail}{\myemail}}}
\end{abstract}

\section{The Wrong Tools for the Job}

Do I need to know the precise polygonal geometries of Los Angeles and the University of Southern California to assert that the latter is within the former? No. My mind contains no such precise geometric model of points and lines, yet I know that USC is in Los Angeles. When humans reason with the real world, they focus on its objects, relations, and processes---rather than starting with geometry---because these are the keys to understanding and explaining the real world. Our GIS tools, however, usually do the opposite. Built from the geometry-up around the legacy logic of traditional cartography (geometries and layers), most GIS tools today are restricted by that legacy's limited ability to model objects, relations, and processes. A representational tension thus exists in GIScience between being a \textit{geometric} information science versus an \textit{ontological}, \textit{relational}, and \textit{processual} information science.

Computational tools help us reason with the world outside. Accordingly, their representations of reality should start with domain theory---well-substantiated systems of ideas to understand and explain phenomena---rather than the constraints of a certain technology or computing platform \citep{gahegan_what_1999,gahegan_our_2018}. For geographic research questions, the relevant domain theory often utilizes object-oriented relations and processes, rather than Cartesian abstractions of space and geometry, even if our computational tools cannot. We need tools that fundamentally embody appropriate scientific theory rather than twisting theory to fit within their representational and computational limitations \citep{harris_more_2017,poorthuis_being_2019}. Although new data sources and knowledge discovery systems can help us wrestle with tricky questions, we impoverish our ability to reason with computers if we do not center theory when we create computational representations of the real world---even if we must rethink or advance our technologies and tools to do so.

It is relatively easy to level such critiques, but if we want better GIS tools to study sociospatial objects, relations, or processes, we need to build them. Urgent tool-building opportunities exist today across many geographic subdomains. As but one interdisciplinary example, consider recent advances in spatially-informed graph models. Modeling spatial dynamics, relations, and topology too often took a backseat historically to geometry, but graphs offer possible ways forward. For instance, geographic knowledge graphs allow us to build spatial information systems around objects and relations\endnote{A simple example of these objects and relations would be: Los Angeles → is a → city; USC → is a → university; USC → is in → Los Angeles} rather than geometries, to better answer ontological spatial questions \citep{yan_spatially_2019}. Yet such tools remain in their infancy today. As another example---and the motivating example on which this paper focuses---graph models of city transportation networks allow topological and dynamical inquiry into urban processes, flows, and structure \citep{barthelemy_spatial_2011,marshall_street_2018}. Yet such tools traditionally relied on network geometry rather than topology (due to data availability and computational constraints), incorporated domain theory poorly, and were usually ad hoc rather than generalizable, accessible, and reusable \citep{boeing_osmnx:_2017,boeing_planarity_2018}.

To conduct better science, we need to build better tools. Such tool-building allows academics to better operationalize and hypothesis-test theory and therefore forms an essential---but poorly incentivized---pillar of scholarly research. In this paper, I reflect on my own tool-building experiences in urban planning and geography: facing the need for a better tool to model and analyze urban street networks in a scalable, theoretically-sound way, I developed a new open-source Python-based software package called OSMnx. This paper considers its history, motivation, and purpose, then reviews its recent use in the empirical street network science literature. In turn, this illustrates the utility of academic tool-building and its downstream---and upstream---value. This paper concludes by proposing better alignment of academic incentives with the positive externalities of conducting open science and developing open-source spatial research software.

\section{If You Want Something Done Right...}

Urban science, sitting at the intersection of city planning, geography, and computational data science, aims to advance our knowledge of cities' fundamental patterns and relationships by modeling spatial big data \citep{batty_new_2013,mattern_methodolatry_2013,solecki_its_2013,kitchin_ethics_2016,sallis_use_2016,alberti_grand_2017,acuto_building_2018,kontokosta_urban_2018,barthelemy_statistical_2019,batty_urban_2019,kang_roundtable_2019,lobo_urban_2020}. Despite urban science's recent bold claims to a \enquote{new kind of science,} urban geographers, sociologists, and planners have of course long investigated cities' patterns and processes through spatial data, mathematical models, and the scientific method \citep{burgess_can_1925,hoyt_is_1951,branch_simulation_1966,batty_modelling_1971,batty_limits_1980,lee_requiem_1973,lee_retrospective_1994,bertuglia_city_1998,johannesen_philosophy_1998,osullivan_physicists_2015,behrend_planning_2019,derudder_engaging_2019}. Computational geography itself now has a long history, yet, too often, geographic science and domain theory fail to fully permeate our computational tools \citep{gahegan_what_1999,harris_more_2017,arribas-bel_geography_2018,gahegan_our_2018,singleton_geographic_2019,gahegan_fourth_2020}.

Why is that? Reflecting on these insufficient links between GIScience---the scholarly field---and GISystems---the software and tools, \citet{gahegan_our_2018} highlights two themes of particular relevance here. First, he argues that the GIScience research community usually does not develop its own software tools because it is in nobody's short-term interests to do so. There are of course exceptions to this rule (some of which will be discussed later) but in general too little geographic science and theory make their way into reusable, accessible tools due to misaligned incentives, expectations, and training in academia. Second, and in turn, Gahegan argues that GIScientists must foster a more robust software development community to build and democratize better scientific research tools that are accessible and available to everyone.

Geography journals have witnessed a recent surge of attention to this under-appreciated importance of academic tool-building. For instance, \citet[][p. 8]{poorthuis_being_2019} argue that \enquote{as a discipline we need to take charge of building and maintaining our own software platforms. These platforms should be open, accessible, and modifiable by the entire academic community and reflect the diversity and heterogeneity of our discipline.} Yet, to date, such tools rarely materialize in practice because, as \citet[][p. 24]{gahegan_our_2018} puts it, we have an \enquote{academic culture that fails to reward those who build or maintain tools and software and encourages a short-sighted and individualistic approach to research.} Along similar lines, \citet[][p. 7]{rey_pysal:_2019} recalls as a junior scholar being told to stop developing tools because \enquote{\enquote{You need to be writing papers.}} He continues, \enquote{My colleagues were being brutally honest and trying to reign in my idealism so that my efforts were more aligned with the realities of promotion and tenure cases at the time.}

The widespread disincentivization of academic tool-building produces several negative outcomes. First, most of our tools rely on impoverished representations of geographic theory because our theoreticians have little incentive or training to build tools. Second, most scientific computational workflows exist only as ad hoc scripts to answer a specific research question before being shelved, rather than being generalizable, documented, shared, and accessible.\endnote{The \enquote{rule of three} in computer programming states that if similar code will be used in three or more places, it should be extracted for general reuse to avoid duplication. We might consider similar guidelines in making our research tools publicly reusable.} An enormous amount of scholarly effort is wasted as we endlessly reinvent each others' wheels. Third and accordingly, reproducibility and replicability remain an outstanding challenge \citep{brunsdon_quantitative_2016,kedron_reproducibility_2019,koster_fueling_2020}. This has become a key motivation for the open-source and open-science movements \citep{rey_show_2009,donoho_50_2017}. But as a scientific community we need to go beyond mere reproducibility and replicability to consider the public \textit{reusability} of our tools and workflows. Otherwise we fail to unlock the broader benefits and spillover effects of tool-building.

These problems plague most geographic disciplines, including the study of cities. Today, a consensus is growing around the importance of harnessing GIScience to the open-source and open-science movements to make urban scientific research more tractable, replicable, theoretical, impactful, and approachable for non-computer scientists. Yet urban science too often lacks open data sources and reusable, accessible, theoretically-sound tools. Incentivizing authors to share their data and computational workflows when submitting journal articles is one nascent step in the right direction. Incentivizing and building reusable, accessible, theoretically-sound tools is another.

\section{Tools for Street Network Science}

Many of these challenges hit close to home for me. A few years ago, I wanted to conduct a nationwide study of US street network form to better understand the fundamental characteristics and outcomes of different urban planning paradigms. How did network structure change as planners reoriented cities around the spatial logic of the automobile? And what do we see in recently-built neighborhoods, considering the rise of neotraditional urban design practice? I was interested in the geometry of these networks, but more important was their topology and the sociospatial dynamics they underlie and organize.

Initially I assumed that some tool must already exist to automatically construct nonplanar directed graph models from ubiquitously available street network topology data. I was wrong on both counts. On the tool side, while some scholars had previously studied similar topics, no one's research software appeared to be publicly available, well-documented for reuse, or sufficiently scalable. Meanwhile, on the data side, mapping platforms like Google Maps did not offer their spatial network data for download. What \textit{was} ubiquitously available was the US Census Bureau's TIGER/Line roads shapefiles, but they place primacy on geometry and contain insufficient topological information to properly model nonplanar networks \citep{boeing_planarity_2018}. I eventually turned to OpenStreetMap.

First launched in 2004, OpenStreetMap is a wiki-style worldwide mapping project and geospatial data repository with good coverage and quality \citep{girres_quality_2010,haklay_how_2010,corcoran_analysing_2013,zielstra_assessing_2013,barron_comprehensive_2014,maier_openstreetmap_2014,basiri_quality_2016,sehra_extending_2019}. To date, over 1 million different users have contributed content including 6 billion nodes (i.e., geospatial points), 600 million ways (i.e., geospatial lines and boundaries), and related descriptive data. It is not a perfect data source: researchers estimate that >95\% of OpenStreetMap contributors are male, suggesting the possibility of correlated biases in content creation \citep{schmidt_gender_2013,graham_towards_2015}. Nevertheless, OpenStreetMap is public, free, and an Open Source Initiative affiliate. Volunteers provide some editorial oversight of edits, but anyone may edit the map using tools such as ESRI's ArcGIS Editor for OpenStreetMap. OpenStreetMap contains data on streets and highways, transit systems, building footprints, parks and plazas, pedestrian and bicycle infrastructure, political boundaries, and more (though non-road coverage varies around the world). But best of all, unlike a shapefile, OpenStreetMap's data model centers spatial objects and their relations, including both geometric and nonplanar topological information.

Researchers typically access OpenStreetMap data through its Overpass API or by downloading a prepackaged regional extract from third-party organizations like Geofabrik. These offer easily ingested street data, but the spatial networks' topological relationships require substantial processing to generate a useful graph model. Researchers often have to write hundreds or thousands of lines of ad hoc code to process the data into a graph and conduct algorithmic analyses for a one-off study. Dozens of small computational and modeling decisions inevitably go unreported in the subsequent peer-reviewed literature, yet these can drastically impact interpretation and replication when every research team codes its own models and analytics. Would it not be better to collectively contribute to and share a reusable, accessible, theoretically-sound set of scientific tools?

One obstacle limiting this tool landscape is that high barriers to entry exist for all but those fluent in computer science and domain theory.  When I first reviewed the urban spatial networks literature, I was struck by how many modeling methodologies made unexplained or even unjustified assumptions about either urban theory or spatial network theory. The landscape lacked tools to handle nonplanar representations of space, which most spatial networks require due to overpasses and underpasses. For simplicity, many studies resorted to undirected graph models, which work fine for studies of form but poorly for studies of flows that obey directionality constraints. The precise handling of common street network features (such as self-loops, parallel edges, or culs-de-sac) often went undocumented. More fundamentally, the Overpass API was cumbersome to work with directly to pipe data into an appropriate model for spatial/network analysis.

In this context I began to develop what eventually became OSMnx. The tool itself is documented\endnote{OSMnx installation instructions and user documentation are available online at \href{https://osmnx.readthedocs.org/}{https://osmnx.readthedocs.org/} and the open-source project itself is hosted at \href{https://github.com/gboeing/osmnx}{https://github.com/gboeing/osmnx}} in detail elsewhere \citep{boeing_osmnx:_2017}, and tutorials and usage examples\endnote{OSMnx tutorials and usage examples in Jupyter notebook format are available online at \href{https://github.com/gboeing/osmnx-examples}{https://github.com/gboeing/osmnx-examples}} are available online, but I briefly summarize its functionality here. OSMnx is a Python package for collecting spatial network data from OpenStreetMap then automatically constructing a directed nonplanar graph model. It is built on top of the open-source geospatial Python stack (which I will return to later). OSMnx's key feature is its ease of use. With just one line of code, the researcher can download and model the street network of any study site in the world: cities, towns, neighborhoods, boroughs, counties, states, nations---any spatial boundary that OpenStreetMap has in its database. The researcher can specify the site's drivable, walkable, or bikeable network. OSMnx includes a suite of visualization tools and graph-theoretic analytics (both geometric and topological) for common transportation planning, urban design, and network science research questions. It can also automatically collect and model elevation, building footprint, and points-of-interest data. The code is documented and open-source, so its formal representations of theory are not black-boxes.

This project began as a few lines of Python code in a Jupyter notebook, before being collected into a module, and later refactored into a formal package distributed online. I am not a software engineer per se, but I knew how to code and was conversant in the relevant domain theory. But, in the midst of working on this project for months on end, I realized that the next person interested in similar empirical questions would face the exact same laborious tool-building process I was then struggling through. Accordingly, I found it useful to make OSMnx open-source for three primary reasons. First, it makes empirical work easier to review and reproduce. Second, it allows anyone else to contribute to the tool's ongoing development. As discussed later, other researchers desiring useful extensions to its functionality have been willing to add them to the codebase. Third, it empowers others working in urban science and planning to advance their empirical research on real-world spatial networks with a reusable, accessible, theoretically-sound tool. I discuss these latter outcomes in the following section.

\section{Empirical Street Network Science with OSMnx}

The teleology of tool-building suggests that the real value lies in the end use of the tool, rather than in its origins. The purpose of developing OSMnx was to conduct empirical research on urban form, travel dynamics, and the topological structure of transportation infrastructure. Since its public release three years ago, several such studies have been conducted---by myself and many others---using OSMnx for model generation, indicator calculation and visualization, and trip simulation. To illustrate the downstream benefits of tool-building, this section briefly reviews the recent empirical literature that uses OSMnx.

Across a variety of academic disciplines and study sites, researchers have recently used OSMnx to download data, generate models, and analyze real-world transportation networks. For example, \citet{hofer_including_2018,hofer_large_2018} model the street network of Graz, Austria to simulate CO\textsubscript{2} emissions and traffic congestion and avoidance behavior using mobility data. \citet{saha_framework_2019} model Mesa, Arizona's streets to generate a synthetic feeder network for electrical distribution. \citet{wang_context-based_2018} model Washington DC's street network to develop a geoprocessing framework for optimizing the meetup locations of multiple people under congested traffic conditions. \citet{liu_spatial_2020} model Beijing's walkable street network to explore spatial patterns of residents' daily leisure activities. \citet{natera_orozco_data-driven_2019} model \enquote{as-is} bicycle networks to demonstrate how cities can make small but targeted infrastructure investments to significantly increase their connectivity and directness. \citet{dumedah_case_2020} model Ghanaian street networks to investigate paratransit service coverage through GPS data. \citet{riascos_networks_2020} model Manhattan's street network for their study of more than one billion taxi trips in New York.

\subsection{Investigating New Urban Technologies}

Studies such as these often investigate the frontier of new transportation and smart cities technologies, including autonomous vehicles, electric vehicles, ride-sharing, and bike-sharing. \citet{beirigo_dual-mode_2018} use OSMnx to model service levels, operational and infrastructure costs, and fleet utilization in hybrid street networks with both autonomous-ready and not autonomous-ready zones. \citet{lin_optimal_2018} model Manhattan's street network alongside travel demand data to optimize ride-share routing. \citet{luo_d3p_2020} model Shanghai's street network to predict demand for electric vehicle sharing systems, while \citet{zhang_electric_2019} model Shanghai's bicycle network to propose a framework for planning dockless bike-sharing services' geofences.

\subsection{Network Structure and Urban Centrality}

Other studies look for fundamental relationships between topological structure---particularly network centrality and robustness---and travel patterns and land use. \citet{wang_road_2020} use OSMnx to model Atlanta's street network to investigate ride-sharing accessibility as a function of network centrality and structure. \citet{wang_street_2018} model Shenzhen's street network to explore the relationship between street centrality and land use intensity. \citet{hellervik_preferential_2019} develop a preferential centrality measure to predict urban activity based on street network structure. \citet{dangelo_modeling_2017} model the street network of Fano, Italy to identify locations of high betweenness centrality. \citet{masias_detecting_2019} operationalize a spatial capture–recapture methodology to model social media users as a function of walkable street network centrality indicators. \citet{dingil_macroscopic_2019} compute and visualize indicators of connectivity, centrality, and clustering across 86 urban areas worldwide to identify the role of network design in easing traffic congestion. \citet{torres_alisis_2019} and \citet{morelli_verificacao_2019} use centrality indicators to measure street networks' vulnerability to perturbation in Mexican and Brazilian cities respectively \citep[see also][]{baumann_assessing_2020,sohouenou_using_2020}.

\subsection{Computer Science Methodological Research}

Computer scientists and statistical physicists often adopt urban transportation networks as tractable, real-world systems that can be well-represented by graphs. OSMnx has been used accordingly to generate input graphs and feature sets for methodological research in machine learning and network algorithms \citep{yin_multi-task_2020,young_automatic_2020}. \citet{ren_deep_2019} model Chengdu's street network then predict traffic flow with a deep spatiotemporal residual neural network. \citet{law_unsupervised_2019} model the street networks at the centers of 100,000 cities worldwide to train a convolutional autoencoder to analyze network structure. \citet{okeeffe_modeling_2019} train a recurrent neural network to reproduce vehicular mobility patterns, using taxi data and street network models. \citet{martinez_mori_bounded_2019} model several cities' street networks to empirically demonstrate heuristic approximation algorithms that make certain network analyses computationally tractable. \citet{feng_spatial_2020} model cities around the world to explore their topology through persistent homology. \citet{shi_optimizing_2018} model Filipino cities to develop a genetic algorithm for optimizing paratransit services in developing countries. \citet{senturk_novel_2019} model Turkish cities to develop a heuristic solution to a clustered variant of the classic traveling salesman problem. \citet{neukart_traffic_2017} model Beijing's street network to develop a quantum annealer for traffic flow optimization on hybrid quantum computing hardware.

\subsection{Indicator Calculation and Visualization}

Other researchers have used OSMnx for automated indicator calculation and visualization. \citet{brandily_roads_2018} calculate street network indicators in 1,800 towns across sub-Saharan Africa to explore the relationship between street density and population growth. \citet{holub_invisible_2017} calculates and visualizes indicators of bicycle network structure and connectivity in Austin, Charlotte, Columbus, and Minneapolis. \citet{quistberg_building_2019} calculate transportation network indicators to build a new platform for conducting cross-country urban health studies. \citet{van_etten_spacenet:_2019} and \citet{verendel_measuring_2019} visualize street network characteristics across their respective studies' sites. \citet{boeing_multi-scale_2020} models the street networks of every US urbanized area, city/town, and Zillow-defined neighborhood to calculate dozens of indicators across tens of thousands of study sites at multiple spatial scales, then shares these models and indicators in a public repository. \citet{wang_impacts_2020} calculate network structure indicators to compare American and Chinese cities. \citet{cherifi_quantifying_2020} compute quality-of-life indicators in Budapest by modeling its pedestrian network and local amenities.

\begin{figure*}[tbh]
	\centering
	\includegraphics[width=1\textwidth]{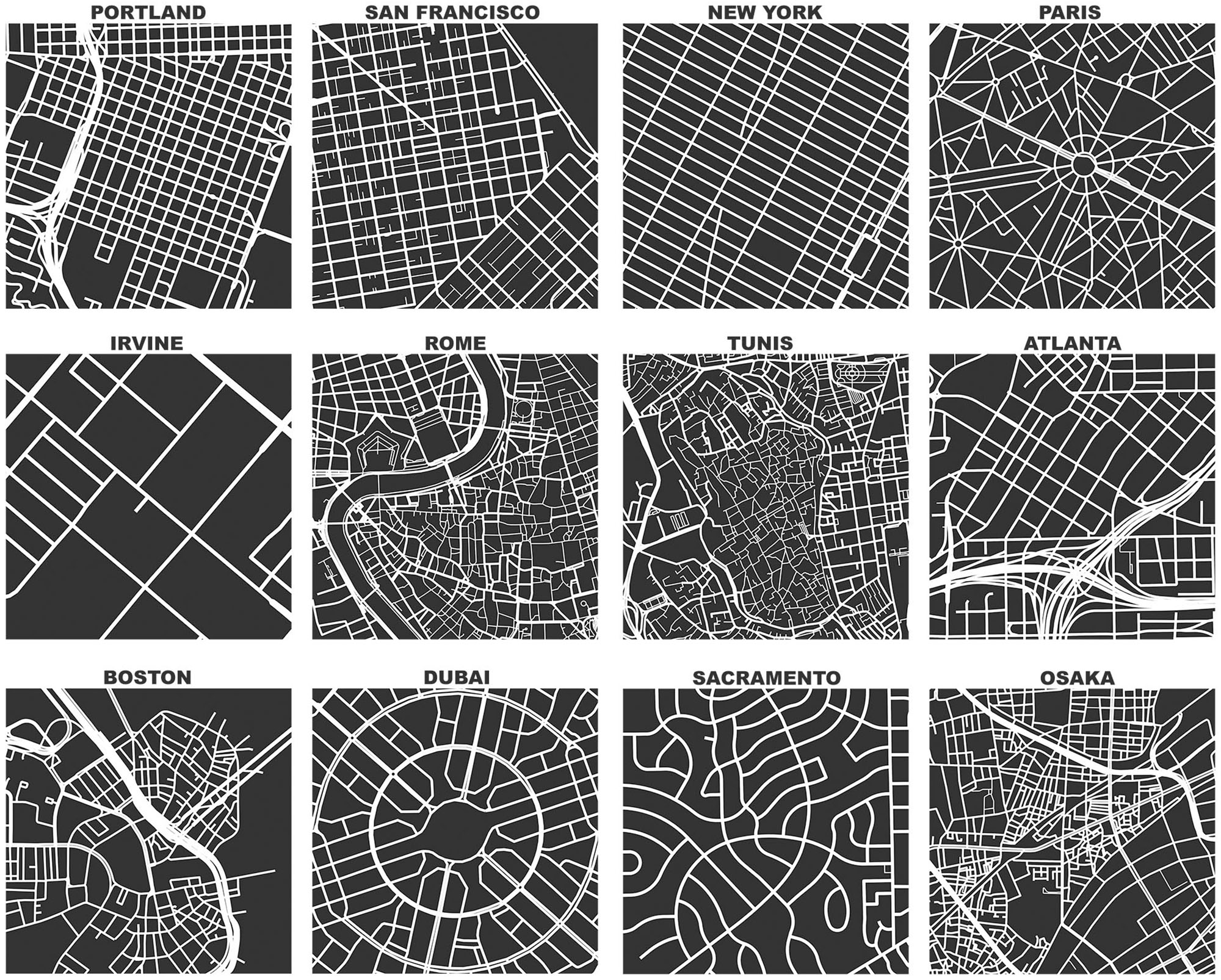}
	\caption{One square mile of different cities' street networks, held at the same scale to compare the urban form and grain \citep{boeing_spatial_2019}.}
	\label{fig:square_miles}
\end{figure*}

\begin{figure*}[tbhp]
	\centering
	\includegraphics[width=1\textwidth]{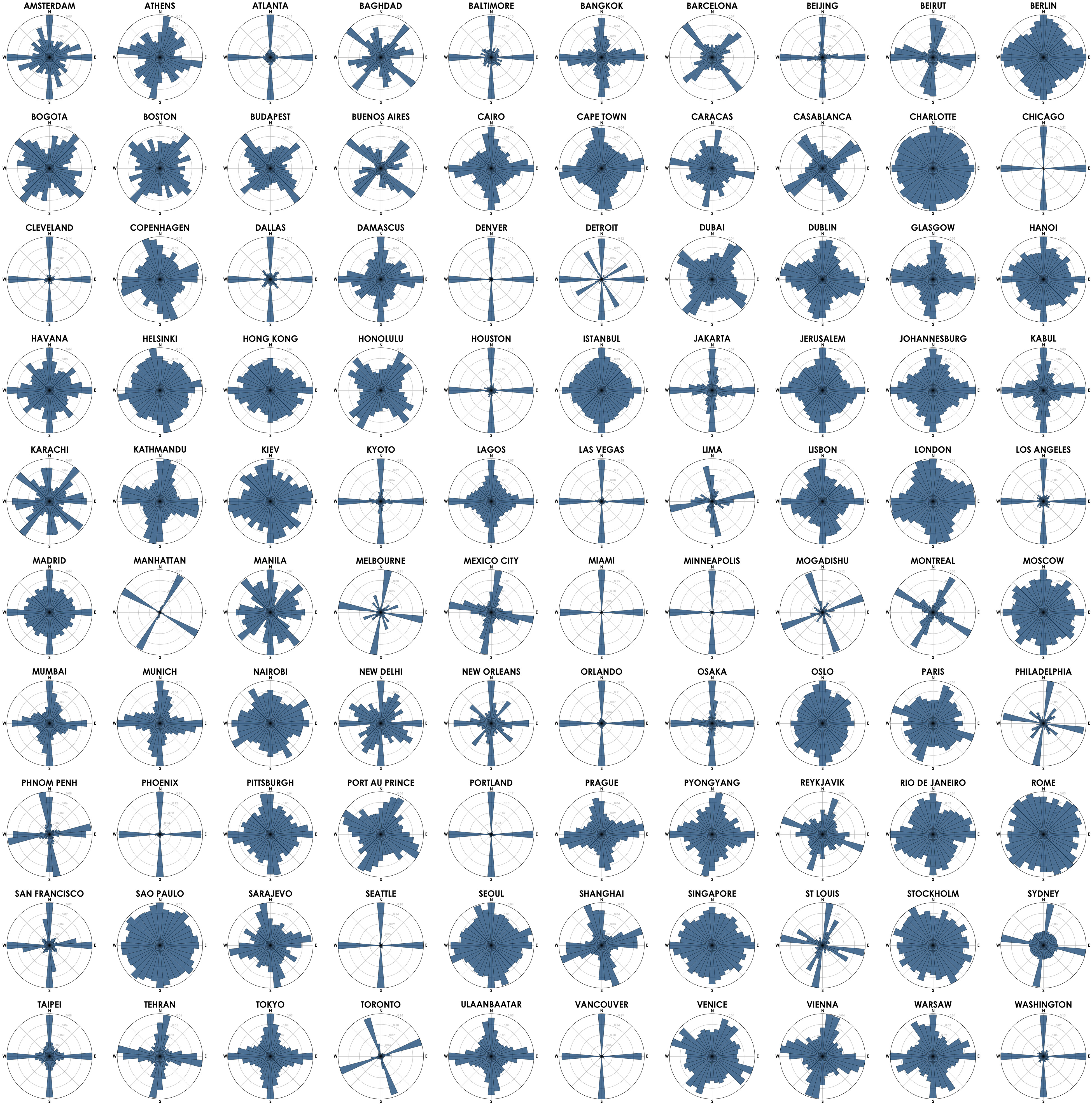}
	\caption{Polar histograms of street orientation in 100 world cities. Each histogram bin represents 10° around the compass, and each bar's length represents the relative frequency of streets with compass bearings falling in that bin \citep{boeing_urban_2019}.}
	\label{fig:street_orientation}
\end{figure*}

\subsection{Urban Morphology}

Several morphological studies of urban form, sprawl, and density have used OSMnx as well. \citet{gervasoni_calculating_2017} measure urban sprawl and network-constrained destination accessibility. \citet{bristow_measuring_2019} compares alternative building density metrics in the context of urban morphology. \citet{abdelkader_topological_2018} and \citet{boeing_planarity_2018} measure urban nonplanarity to retheorize how planar assumptions impact street network analyses and to suggest better models. The World Bank collects building footprint data then trains a random forest model to identify OpenStreetMap coverage gaps that can inform future crowd-sourcing efforts and mapping campaigns \citep{jones_identifying_2019}. \citet{boeing_spatial_2019} explores the growing role of big data in computational urban morphology and visual analytics (Figure \ref{fig:square_miles}). \citet{boeing_urban_2019} conducts a cluster analysis of urban street networks around the world to theorize how planners produce different forms of geometric spatial order, illustrated in Figure \ref{fig:street_orientation}. \citet{coutrot_cities_2020} build on this theory to explore how spatial order impacts a city's residents' spatial navigation, finding that residents of more grid-like places demonstrate worse ability.

\subsection{Network-Constrained Trip Simulation}

Finally, several studies have used OSMnx for simulating trips along a network. \citet{waddell_architecture_2018} build regional planning models that integrate land use simulation, travel demand modeling, and traffic assignment. \citet{hernandez-hernandez_anger_2019} calculate commute routes and trip distances in a study of motorists' emotions and expressions of anger in Mexico City. \citet{boeing_morphology_2019} compares driving versus walking route circuity by simulating millions of trips across 40 city street networks. \citet{merchan_quantifying_2020} and \citet{merchan_empirical_2019} investigate last-mile logistics circuity in São Paulo and develop better approximation algorithms for urban route distance prediction. \citet{padgham_introduction_2019} examine hospital siting by simulating network-constrained stroke service center catchment basins, while \citet{lin_analysis_2019} identify the determinants of bicycle catchment basins around Shanghai's metro stations. \citet{liao_disparities_2020} model street networks in several world cities to compare travel time by personal automobile versus public transit. Finally, several recent studies have used OSMnx during the global COVID-19 pandemic for spatial epidemiology and healthcare accessibility modeling \citep[e.g.,][]{adler_geospatially_2020,kang_rapidly_2020}.

\subsection{Summary}

The preceding survey of recent studies covers a wide range of spatial topics and disciplines, from urban design to public health to transport engineering to computer science. These research projects investigate various travel modes, from walking to cycling to driving to ride-hailing. They study neighborhoods and cities in every inhabited continent. Most importantly, I could not have conducted most of these studies myself as they exceed my knowledge and skill, but by making OSMnx open-source and accessible, it has percolated into others' research designs around the world. Academic tool-building thus entails upstream and downstream benefits for our field's wider endeavor of scientific discovery and theory building.

\section{Tool-Building in Academia}

A couple of years ago, I gave a talk which partly dove into OSMnx development and the benefits of academic tool-building and open science. At dinner that evening a senior faculty member asked if my development work continued. I said yes and listed a few features in development that would unlock exciting new spatial network analytics and research questions. The faculty member pushed back on this and---echoing \citet{rey_pysal:_2019}'s recollections as a junior scholar---suggested that if I were to continue in academia, that I must give up tool-building. To paraphrase: \enquote{You will become known merely as a tool builder rather than a serious scholar. A serious scholar cannot waste time on anything but empirical research and advancing theory.}

Must scholars eschew tool-building in order to successfully further their field? I believe, rather, that we impoverish our field if we do not make our otherwise ad hoc, one-off research tools \enquote{open, accessible, and modifiable,} to return to Poorthuis and Zook's earlier words. In particular, there are three primary benefits of open-source tool-building for academics: 1) unlocking your own empirical research, 2) advancing the collective scientific and theoretical endeavor, and 3) impacting a broader audience.

\subsection{Unlocking Individual Research}

First, tool-building unlocks your own research. Nearly all of us in the spatial sciences do \textit{some} tool-building as we create simple macros, scripts, or libraries of code for routinizing mundane data processing tasks or fitting models or generating graphical and tabular output for subsequent publication. In this era of big data and ubiquitous computation, it is inefficient and limiting to rely on point-and-click interfaces to conduct scientific research. Coding skills thus increasingly appear in spatial science curricula as necessary components of any GIS skillset. Fostering such abilities is crucial as data science and coding grow central to spatial analysis. But most importantly, as researchers motivated to ask new questions and develop new methods, we cannot assume that theory-rich tools already exist to answer today's important questions. We must build the right tools to answer the right questions---and doing so can open up new personal research trajectories.

\subsection{Advancing the Discipline}

Second, once we have built these tools, they should be open-source and accessible to advance the wider discipline. Fortunately this philosophy already has traction in the GIScience and urban science communities. Efforts are underway today to build a national geospatial software center \citep{geospatial_software_institute_gsi_2020} and other examples abound. Luc Anselin's GeoDa and GeoDaSpace software for spatial analysis and econometric modeling are free and open-source \citep{anselin_geoda:_2006}, as is Serge Rey's Python Spatial Analysis Library (PySAL) \citep{rey_open_2015,rey_pysal:_2019}. Paul Waddell's UrbanSim platform, a Python-based software library that supports urban land use and transportation modeling and simulation, is similarly free and open-source \citep{waddell_urbansim:_2002}. Beyond this sort of \textit{hero model} of tool-building, many other community-driven open-source spatial software projects exist in the Python ecosystem and other programming languages (e.g., GDAL, geopandas, leaflet, PostGIS, QGIS, r-spatial, and even OpenStreetMap itself). Such projects often utilize a decentralized \textit{many hands model} to grow a community around building better tools and collaborating synergistically around shared goals. Regardless of the exact model, academic-led projects can provide a theory-rich scaffolding on which research can be conducted and other tools (like OSMnx) can be built. These tools and their ongoing development are also important for empowering students and scholars who cannot afford expensive proprietary software licenses. We must reduce such barriers to entry.

Academics too often pull up the gangplanks by placing scientific findings behind paywalls, hoarding useful datasets, or concealing their research software. The intertwined open-science, open-data, and open-source movements address these three respectively by publicly sharing scientific findings, data, and software for the good of society. This in turn disseminates knowledge and empowers the wider scientific community. If our goal as academics is to produce empirical research and advance theory, our time \textit{should} focus on research and writing---but we should also set aside time to build better theory-rich tools to answer difficult questions.

\subsection{Broader Community Impacts}

Third, open-source tool-building impacts a broad community with bidirectional effects as the tools we build eventually underpin others' work downstream, while other researchers contribute back to our projects upstream. As two examples of downstream effects, the mobile crowd-sensing platform CrowdSenSim now uses OSMnx to simulate urban environments \citep{tomasoni_profiling_2018,montori_crowdsensim_2019} and the transportation planning company Remix developed its platform---now deployed in hundreds of cities worldwide---initially using OSMnx to model street networks. Reciprocally, regarding the upstream direction, as an open-source project OSMnx has received hundreds of code contributions from other scholars and members of the public. These contributors helped develop its points-of-interest module, its nearest-node and nearest-edge search algorithms, its building footprint functionality, and much more. I have reaped the benefits of dozens of others' contributions that enhanced the tool in ways I subsequently used to answer my own research questions.

Better tools and data models, spearheaded by academics, can help infuse theory into our field's quantitative work. But if we want better tools, we have to build them. It is not ESRI's job to satisfy all the theoretical needs of the spatial sciences. Recent important scholarly critiques have highlighted how today's GIScience tool landscape is inaccessible, atheoretical, and ad hoc \citep[e.g.,][]{gahegan_our_2018,poorthuis_being_2019}. One clear path to better link this academic critique with tangible real-world action is to build and incentivize better tools for praxis. Collectively, we need to spend more time fixing the lack of high-quality accessible tools rather than just writing yet another thinkpiece lamenting their dearth.

\section{Toward Better Tools}

Software often feels inevitable because its backstory is often invisible. We click a download link, run an installer, and suddenly have a new tool to use. Yet this conceals years of human decisions, experiences, and constraints shaping software outcomes that are in no way predestined. In many ways, open-source software parallels other public infrastructure such as highways and bridges. Humans plan and engineer infrastructure in specific social, economic, and technological contexts. Like an individual highway's or bridge's broader network of connected roads, any single piece of software represents a complex network of entanglements with many other software projects on which it depends. And like a highway or bridge, software requires years of maintenance, updates, and retrofits after its initial development: although splashy new capital projects often receive funding and adulation, critical routine maintenance work usually receives far less of either.

Tool-building, in all its facets, remains an essential but poorly incentivized pillar of academia today. Computational geographic tools too often impoverish quantitative analytics through poor representations of scientific theory and squander precious time as everyone develops their own ad hoc scripts to solve similar problems. The case of urban street network science illustrated this. Building an open-source, reusable, accessible, theoretically-sound tool as public infrastructure has generated various downstream and upstream benefits. So what can we do to foster a robust tool-building community and collectively reap more of these benefits?

Incentives are key. First, academic tenure, promotion, and annual review guidelines should explicitly reward the scholarly value of open-source research software and open data contributions to better acknowledge their significant value. They should also account for contributions to pre-existing and decentralized projects to encourage collaborative progress and maintenance of the open-source commons. An ideal system would better balance the respective value of research, publication, and tool-building to advancing science and theory. Second, we should train the next generation of practitioners and scholars to be better tool creators and consumers. Curricula should include coding and informatics courses and pedagogy should emphasize hands-on learning, such as using computational notebooks. Third, academic publication must continue its nascent steps toward open science: journal editors should require the submission and publication of datasets and computational workflows alongside quantitative manuscripts. Finally, we should increase funding opportunities for building and maintaining research software to foster the positive externalities and downstream benefits they generate throughout the research community.

The goal of academic tool-building---and the hope of every such tool-builder---is to construct some kind of useful infrastructure for your field. With better scaffolding in place to connect theory, science, and analytics, we can conduct better research to explore important geospatial questions and foreground objects, relations, and processes. OSMnx has not changed the world or reinvented urban science, but it has made empirical street network analysis a little easier and more reproducible for some urban scholars and practitioners. Hopefully it even unlocked a study or two along the way that otherwise could not have been conducted. Finally, there is one key lesson-learned that I would like to share with other budding academic tool-builders: give your tool an easier name to remember and pronounce than OSMnx.

\section*{Acknowledgments}

This work was funded in part by a grant from The Public Good Projects. I wish to thank the editors of \textit{Transactions in GIS} for the opportunity to present this plenary paper and two anonymous reviewers for their helpful suggestions in improving it. I also wish to thank the innumerable other developers of open-source software on which my research and tool-building rely. Figure \ref{fig:square_miles} is reprinted with permission from \citet{boeing_morphology_2019} and Figure \ref{fig:street_orientation} is reprinted from \citet{boeing_urban_2019} under the terms of the \href{http://creativecommons.org/licenses/by/4.0/}{Creative Commons Attribution 4.0 International License}.

\bigskip
\IfFileExists{\jobname.ent}{\theendnotes}{}
\bigskip

\RaggedRight
\setlength{\bibsep}{0.00cm plus 0.05cm} 
\bibliographystyle{apacite}
\bibliography{TGIS-plenary}

\end{document}